\documentstyle[preprint,aps,epsfig]{revtex}
\begin{document}
\preprint{ CERN-TH/2000-273}
\title{Testing LSND at long-baseline neutrino experiments}
\author 
{Gabriela Barenboim}
\address
{Theory Division, CERN, CH-1211 Geneva, Switzerland} 
\author
{Mohan Narayan and S. Uma Sankar}  
\address
{Department of Physics, I.I.T., Powai, Mumbai 400076, India}
\date{\today}
\maketitle

\begin{abstract}
Recently it was suggested that two very different mass-squared
differences play a role in atmospheric neutrino oscillations.
The larger of these also accounts for the LSND result and the
smaller of these also drives the solar neutrino oscillations.
We consider the predictions of this scheme for long-baseline
experiments. We find that high statistics experiments, such as
MINOS, can observe a clean signal for this scheme, which is
clearly distinguishable from the usual scheme of atmospheric
neutrino oscillations driven by a single mass-squared difference.
\end{abstract} 
\vspace{0.5cm}

\section{Introduction}
At present there are three pieces of evidence for neutrino flavour conversion:
\begin{enumerate}
\item \underline{Solar Neutrino Problem} The measured flux of $\nu_e$  
from the Sun is smaller than the expected flux \cite{nu2k-sol}.
\item \underline{Atmospheric Neutrino Problem} The measured flux of 
$\nu_\mu$, generated by the cosmic ray interactions in the atmosphere, is 
smaller than the Monte Carlo expectation \cite{nu2k-atm}.
\item \underline{LSND} The LSND experiment has observed signals for
both $\bar{\nu}_\mu \rightarrow \bar{\nu}_e$ and $\nu_\mu \rightarrow
\nu_e$ transitions \cite{nu2k-lsnd}.  
\end{enumerate}
If each set of data is analysed under the assumption that only two 
neutrino flavours oscillate into each other, then the following 
constraints are obtained:
\begin{eqnarray}
10^{-6} {\rm~eV}^2 < \Delta m^2_{sol} < 10^{-4} {\rm~eV}^2, &~~~~~& 
\theta_{sol} \sim 3^\circ ~{\rm or}~ \sim 30^\circ \nonumber \\
10^{-3} {\rm~eV}^2 < \Delta m^2_{atm} < 10^{-2} {\rm~eV}^2, &~~~~~& 
\theta_{atm} \sim 45^\circ \nonumber \\
0.2 {\rm~eV}^2 < \Delta m^2_{LSND} < 0.4 {\rm~eV}^2, &~~~~~& 
0.01 < \sin^2 2 \theta_{LSND} < 0.1. \label{eq:param}
\end{eqnarray} 
The data from the Bugey accelerator provide the lower limit on $\Delta 
m^2_{LSND}$ \cite{bugey} and the CDHSW data provide the upper limit
\cite{cdhsw}. 

From the constraints on various $\Delta$'s, it seems as if one may not  
be able to account for all the positive results in the framework of
three-flavour neutrino oscillations. Very often, three-flavour oscillation
fits are done using solar and atmospheric neutrino data only. In this
scheme, which we call the standard scheme, there are two independent 
mass-squared differences and three mixing angles. The solar neutrino  
oscillations depend on only the smaller mass-squared difference (which is
set equal to $\Delta m^2_{sol}$) and two mixing angles $\theta_{12}$ and
$\theta_{13}$ \cite{kuopan}. The atmospheric neutrino oscillations
depend on the larger mass-squared difference (which is set equal to 
$\Delta m^2_{atm}$) and two mixing angles $\theta_{13}$ and $\theta_{23}$
\cite{jim94}. The CHOOZ experiment constrain $\theta_{13}$ to be very small
($< 9^\circ$) \cite{chooz,nru}. Since the common parameter between solar 
and atmospheric neutrino oscillations is very small, these two oscillations 
effectively become two different two-flavour oscillations. Hence, the
above constraints on $\theta_{sol}$ and $\theta_{atm}$ apply directly  to 
$\theta_{12}$ and $\theta_{23}$, respectively. Thus we find that, in this  
scheme, the preferred solution to the atmospheric neutrino problem is  
$\nu_\mu \rightarrow \nu_\tau$ oscillations with maximal mixing.
 
The long-baseline experiments are designed to test the hypothesis that
$\nu_\mu \rightarrow \nu_\tau$ oscillations, with $\Delta m^2_{atm} 
\simeq 3 \times 10^{-3}$ eV$^2$, are the cause of the atmospheric
neutrino deficit. K2K and MINOS will look for muon neutrino disappearance.
The number of $\nu_\mu$ charged current (CC) events in these experiments
is given by the convolution of their neutrino spectrum with the
$\nu_\mu$ survival probability: 
\begin{equation}
P_{\mu \mu} = 1 - \sin^2 2 \theta \sin^2 \left( 1.27 
\frac{\Delta m^2 L}{E} \right),
\end{equation}
where $\Delta m^2$ is in eV$^2$, $L$ is in km and $E$ is in GeV.
In most neutrino oscillation experiments, there is a single constraint
involving two unknowns, $\theta$ and $\Delta m^2$. We can constrain one 
of them, only by making an assumption about the other parameter. However,
MINOS is a very high statistics experiment and can measure the spectrum of
$\nu_\mu$ CC events. This spectrum will have a minimum at $E = E_{min}$,
where $(1.27 \Delta m^2 L/E_{min}) = \pi/2$. The number of events at this
minimum is proportional to $(1-\sin^2 2 \theta)$. Hence MINOS can
determine the mass-squared difference and the mixing angle independently.
If the standard scenario is correct, then $\sin^2 2 \theta_{atm} =1$ and 
the number of events at $E = E_{min}$ should be zero. The energy of the 
neutrino beam for MINOS can be tuned in such a way that the minimum
will occur where the beam flux is substantial. Hence MINOS is sensitive 
to the whole range of $\Delta m^2_{atm}$ suggested by the atmospheric 
neutrino data. 

To account for the three signals for oscillations, it seems as if 
three-flavour oscillations are inadequate and one must introduce at 
least one more light neutrino. Since the measurement of the invisible 
width of $Z^0$ boson shows that there are only three light active 
neutrinos, the forth neutrino must be sterile. Four-flavour oscillations 
between three active and one sterile neutrino, with three independent 
mass-squared differences set equal to the above three scales,
are considered extensively \cite{bggrev}.

Since no direct evidence for any sterile neutrino has been seen, then
it is worth re-examining the simple assumption that only a single 
$\Delta$ plays a role in each of the above evidences for oscillations.
Recently it was suggested by Scheck and Barenboim (SB) that oscillations 
between three active flavours may be able to account for all three 
signals \cite{barsch}. In this scheme it is assumed that the larger 
mass-squared difference, $\Delta_{32}$, is about $0.3$ eV$^2$ and 
drives the LSND oscillations. The smaller mass-squared difference
$\Delta_{21}$ is assumed to be small enough that LSND is not sensitive
to it. The key assumption in this scheme is that both  $\Delta$'s
play a role in creating the deficit of $\nu_\mu$ flux in the atmospheric
neutrino problem. $\Delta_{32}$ drives the oscillations of the 
downward going neutrinos. Since the path length of these neutrinos
is small, these oscillations are not sensitive to $\Delta_{21}$. 
The magnitude of $\Delta_{21}$ is fixed by two requirements: 
a) the zenith-angle dependence of the deficit of upward going $\nu_\mu$'s 
should be reproduced and b) the solar neutrino deficit should be 
adequately explained. These requirements fix $\Delta_{21}$ to be in the
range $10^{-4}$--$10^{-3}$ eV$^2$. This value of $\Delta_{21}$ is much
larger than $\Delta m^2_{sol}$ given in Equation~(\ref{eq:param}). 
However, the latest Super-Kamiokande data on solar neutrinos do prefer 
$\Delta m^2_{sol} \simeq 10^{-4}$ eV$^2$ \cite{nu2k-sol}.

In this letter, we consider the signals that will be observed at K2K 
and MINOS in the SB scheme. We find that at K2K the signals in the SB
scheme are somewhat different from those in the standard scheme. With 
an accumulation of events over a period of time, it may be possible
to differentiate between the two schemes using K2K data. However for
MINOS, the signals predicted by SB scheme for the $\nu_\mu$ CC event
spectrum are qualitatively different from those in the standard scenario. 
Moreover, the oscillations due to the $\Delta_{32} \simeq 0.3$ eV$^2$
should be clearly visible if MINOS runs in its high energy beam mode. 

\section{Scheck-Barenboim (SB) Scheme}

Let us briefly recall the salient features of the SB scheme. The three
active flavours mix to form three mass eigenstates
\begin{equation}
| \nu_\alpha \rangle = U |\nu_i \rangle,
\end{equation}
where $U$ is a $3 \times 3$ mixing matrix parametrized by three mixing   
angles and a CP-violating phase. Here for simplicity we set the phase to
zero. The matrix $U$ can be written in the form
\begin{equation}
U = U_{23} (\theta_{23}) U_{13} (\theta_{13}) U_{12} (\theta_{12}).
\end{equation}
Without loss of generality, we can assume that the masses satisfy the 
inequalities $m_1 < m_2 < m_3$. Then the mixing angles should have
the range $(0, \pi/2)$ to cover all the possibilities that are 
physically distinguishable. The independent mass-squared differences
are taken to be $\Delta_{21} = m_2^2 - m_1^2$ and $\Delta_{32} =
m_3^2 - m_2^2$. It is assumed that $\Delta_{32} \simeq 0.3$ eV$^2$ 
to account for the LSND results, and the magnitude of $\Delta_{21}$ is 
taken to be in the range $10^{-4}$--$10^{-3}$ eV$^2$ so that the 
zenith-angle dependence of atmospheric neutrinos and solar neutrino
suppression are reproduced. 

In Ref. \cite{barsch} the ranges of the mixing angles allowed by solar,
atmospheric and LSND data were obtained, with values of $\Delta$'s
as given above. We have updated their results by including the further
constraints from Bugey \cite{bugey}, CHOOZ \cite{chooz}, CHORUS 
\cite{chorus} and NOMAD \cite{nomad}. CHORUS and NOMAD have very 
small values of $(L/E)$ and hence they give no meaningful constraints
on mixing angles for the values of $\Delta$'s we consider here. For 
Bugey the average value of $(L/E)$ is about 11 and it is sensitive
to $\Delta_{32}$ but not to $\Delta_{21}$. The oscillations driven by 
$\Delta_{32}$ are averaged out at CHOOZ because it has $\langle L/E \rangle 
\sim 300$. CHOOZ is sensitive to $\Delta_{21}$ if it is as large as 
$10^{-3}$ eV$^2$, but it is insensitive to smaller values. Both Bugey and
CHOOZ require $\theta_{13}$ to be small and together they yield the
constraint 
\begin{equation}
\theta_{13} \leq 9^\circ. \label{eq:13lim}
\end{equation}
Depending on the value of $\Delta_{21}$ CHOOZ also constrains 
$\theta_{12}$. If $\Delta_{21} \simeq 10^{-3}$ eV$^2$, then
we get the constraint $\theta_{12} \leq 10^\circ$. However, if  
$\Delta_{21} \leq 7 \times 10^{-4}$ eV$^2$ then $\theta_{12}$ is 
unconstrained. 

The $\nu_\mu \rightarrow \nu_e$ oscillation probability relevant to LSND
in this scheme is given by
\begin{equation}
P_{\mu e} = \sin^2 \theta_{23} \sin^2 2 \theta_{13} \sin^2 
\left( 1.27 \frac{\Delta_{32} L}{E} \right).
\end{equation}
We need both $\theta_{13}$ and $\theta_{23}$ to be non-zero to explain 
the positive signal of LSND. The allowed range of $\theta_{23}$ is a 
function of $\theta_{13}$. The smallest allowed value of $\theta_{23}$,
which will be relevant to long-baseline experiments, occurs for the
largest value of $\theta_{13} = 9^\circ$. For this value, we have 
\begin{equation}
20^\circ \leq \theta_{23} \leq 50^\circ. 
\end{equation} 
In Ref. \cite{barsch}, it was shown that, to explain the atmospheric 
neutrino problem, one needs $\theta_{23} \simeq 27^\circ$, which is well 
within the above range.

\section{Signals at long-baseline experiments}

The $\nu_\mu$ survival probability for the case of three active 
flavour oscillations is given by  
\begin{eqnarray}
P_{\mu \mu} & = & U_{\mu 1}^4 + U_{\mu 2}^4 + U_{\mu 3}^4 + 
2 U_{\mu 1}^2 U_{\mu 2}^2 \cos \left(2.53 \frac{\Delta_{21}L}{E} \right)+ 
\nonumber \\ & & 
2 U_{\mu 1}^2 U_{\mu 3}^2 \cos \left(2.53 \frac{\Delta_{31}L}{E} \right)+ 
2 U_{\mu 2}^2 U_{\mu 3}^2 \cos \left(2.53 \frac{\Delta_{32}L}{E} \right). 
\end{eqnarray}
The K2K experiment has a baseline length of 250 km and its neutrino energy 
spectrum is peaked around 1 GeV \cite{oyama}, so it has an $(L/E)$ value 
of about 250. For this large a value of $(L/E)$, the oscillations due
to $\Delta_{32}$ of the SB scheme get averaged out. K2K is not sensitive to 
values of mass-squared differences smaller than $10^{-3}$ eV$^2$, 
hence $\Delta_{21}$ of the SB scheme can be set to zero. Under these
approximations, the $\nu_\mu$ survival probability relevant to K2K is 
\begin{equation}
P^{K2K}_{\mu \mu} = (U_{\mu 1}^2+U_{\mu 2}^2)^2 + U_{\mu 3}^4
= (1 - U_{\mu 3}^2)^2 + U_{\mu 3}^4.
\end{equation}
Here, $U_{\mu 3} = \sin \theta_{23} \cos \theta_{13} \simeq \sin \theta_{23}$
because $\theta_{13}$ is constrained to be small. For $\theta_{23} \simeq
27^\circ$, we have $P^{K2K}_{\mu \mu} = 0.67$. The expected number of 
$\nu_\mu$ CC events at K2K is obtained by convoluting $P_{\mu \mu}$ with
the energy spectrum of the neutrino beam; the integral of the spectrum gives 
the expected number of events in case of no oscillations. The ratio of the 
above two numbers is purely a function of the oscillation parameters.
We saw above that in the SB scheme $P_{\mu \mu}$ for K2K is independent 
of energy. Hence the ratio of expected number of events with and without 
oscillations is equal to the constant $P_{\mu \mu} = 0.67$. This number,  
predicted by the SB scheme, is to be contrasted with $0.46$, which is the  
prediction of the standard scheme, in which the atmospheric neutrino
problem is assumed to be due to $\nu_\mu \rightarrow \nu_\tau$ oscillations
with maximal mixing and $\Delta m^2_{atm} \simeq 3 \times 10^{-3}$ eV$^2$  
and $\Delta m^2_{sol} \ll \Delta m^2_{atm}$. The prediction of the standard  
scheme rises to $0.8$, if one takes $\Delta m^2_{atm} \simeq 10^{-3}$ eV$^2$, 
which is the smallest value allowed by Super-Kamiokande data in that
scheme. Thus K2K data may rule out the SB scheme if the measured
suppression turns out be less than $0.5$. For larger values of the 
suppression, however, the K2K data will not be able to distinguish
between the two schemes. 

Because of the limited statistics in K2K, one measures only the total
number of $\nu_\mu$ CC events but not their spectrum. Because of this,
the K2K results will not be able to provide an unambiguous signal for the 
two $\Delta$ solutions to the atmospheric neutrino problem. MINOS, being a 
high statistics experiment, measures the spectrum of $\nu_\mu$ CC 
events. This measurement allows them to determine the mass-squared  
difference and mixing angle independently if only one mass-squared 
difference plays a role in atmospheric neutrino oscillations. The
same measurement will also enable them to determine if two mass-squared
differences play a role in atmospheric neutrino oscillations. 

MINOS has a baseline length of 730 km and it has three options for 
the energy of its neutrino beam. For the low energy option, the energy
is peaked around 3 GeV, which corresponds to $(L/E) \sim 250$. For 
the medium energy option, $E_{peak} = 7$ GeV, which means $(L/E) \sim
100$. For the high energy option, $E_{peak} = 15$ GeV with $\langle L/E 
\rangle \simeq 50$. We obtained the spectra for these three options from
Ref. \cite{mintdr} and multiplied them with $P_{\mu \mu}$ to obtain the
energy distribution of $\nu_\mu$ CC events. The distributions for the low
energy option are plotted in Fig.~1, where the thick line is the prediction 
of the SB scheme, the dotted line is the expected spectrum in case of no 
oscillations, and the thin line is the prediction of the standard scheme with 
$\Delta m^2_{atm} = 3 \times 10^{-3}$ eV$^2$ and $\theta_{atm} =
45^\circ$. As mentioned in the introduction, the event distribution
touches zero at the point where the energy satisfies the equation $1.27
\Delta m^2_{atm} 730/E = \pi/2$. This is an unavoidable feature of
atmospheric neutrino oscillations with one mass-squared difference, because 
in such a scenario the corresponding mixing angle is constrained to be
maximal. However, we see that in the SB scheme, where two $\Delta$'s play
a role in atmospheric neutrino oscillations, the event distribution will
not touch zero because of the interplay between the two types of
oscillations. Between 1 and 5 GeV, the smallest suppression in any energy
bin is about $0.5$. Such a signal can give a striking proof regarding the
role of two $\Delta$'s in atmospheric neutrino oscillations. The predictions 
for the SB scheme are plotted for $\Delta_{32} = 0.3$ eV$^2$, $\Delta_{21}
= 3 \times 10^{-4}$ eV$^2$, $\theta_{12} = 35^\circ$, $\theta_{13} =
9^\circ$ and $\theta_{23} = 27^\circ$. Changing the value of $\Delta_{21}$
in the allowed range does not qualitatively affect the form of the
signal. For neutrino energies  above 10 GeV, one clearly sees oscillations
of $0.3$ eV$^2$ generated by $\Delta_{32}$. This signal can be more
clearly seen in the high energy neutrino beam of MINOS. In Fig.~2 we
plotted the prediction of the SB scheme (thick line) along with the
expectation in the case of no oscillations. The oscillations are more
clearly visible in this case. This may be the first instance of the
observation of a variation of oscillations as a function of energy. 

In plotting Figs. 1 and 2, we assumed that the energy resolution of
MINOS is very good, better than $0.5$ GeV or so. Therefore the number of
events is not smeared with the energy resolution. However, the signal
for the SB scheme is markedly different from that of the standard scheme, 
even if the energy resolution is worse than $1$ GeV. Then in the low
energy option, the suppression seen will be about $0.6$ for the SB scheme
whereas it will be about $0.3$ for the standard scheme. But it is in the
high energy option that the predictions of the SB and standard schemes 
are qualitatively different. Here the standard scheme predicts no 
suppression at all for the entire allowed range of $\Delta m^2_{atm}$. 
The SB scheme predicts a minimum suppression of about $0.6$ in the high 
energy option, for all the allowed values of the parameters. Hence even  
with bad energy resolution, MINOS is capable of distinguishing between 
the standard scheme and the SB scheme.

It was  mentioned in the introduction that four-flavour oscillations 
(three active and one sterile) are considered to account for LSND results. 
In these schemes each of the pieces of evidence for flavour conversion
of neutrinos (solar, atmospheric and LSND) is explained by oscillations 
driven by their own individual $\Delta$. The various types of
four-flavour oscillations are summarized in Ref. \cite{bggrev}.
The combined data restrict the solar neutrino oscillations to be 
essentially $\nu_e \rightarrow \nu_s$ (where $\nu_s$ is the sterile 
neutrino) and the atmospheric neutrino oscillations to be essentially 
$\nu_\mu \rightarrow \nu_\tau$ oscillations. We calculated the predictions
of four-flavour oscillations for MINOS for the following values of 
the parameters:
\begin{itemize}
\item 
$\Delta m^2_{LSND} = 0.3$ eV$^2$ 
and $\sin^2 2 \theta_{LSND} = 0.1$
\item 
$\Delta m^2_{atm} = 3 \times 10^{-3}$ eV$^2$ 
and $\sin^2 2 \theta_{atm} = 1$
\end{itemize}
The long-baseline experiments are not sensitive to $\Delta m^2_{sol}$.
The results we obtained are indistinguishable from those of the 
standard scheme for both low and high energy beams of MINOS. This 
occurs for the following reason. The mass-squared difference
to which the long-baseline experiments are the most sensitive is 
$\Delta m^2_{atm}$. In both the standard scheme and the four-flavour 
scheme, it has the same value. Hence $P_{\mu \mu}$ in both schemes is
very similar. In the four-flavour scheme, the larger mass-squared 
difference $\Delta m^2_{LSND}$ gives rise to some modification of
$P_{\mu \mu}$, but these modifications are small because $\theta_{LSND}$
is small. This is illustrated in Fig.~3.

\section{Conclusion}

We considered the signals at the long-baseline experiments K2K and MINOS
as predicted by two different mixing schemes of three active flavours.
In the standard scheme only one $\Delta$ is assumed to drive  
atmospheric neutrino oscillations, while the other $\Delta$ is much 
smaller. In the SB scheme both $\Delta$'s play a role in 
atmospheric neutrino oscillations and the larger $\Delta$ also 
drives LSND oscillations. The K2K experiment may be able to distinguish
between these two schemes for some values of the allowed parameters. 
However, the energy distribution of the events at MINOS, in both low and
high beam energy options, can provide a clear distinction between the 
two scenarios for any of the allowed values of the parameters.

We also considered the signals at MINOS from four-flavour oscillations.
These signals are indistinguishable from those of the standard scheme. 
However the BooNE experiment at Fermilab \cite{nu2k-bne} can verify or 
rule out LSND results. If BooNE confirms LSND results, then the
high energy option of MINOS should be pursued. The results of MINOS 
will tell us whether LSND results can be incorporated within three active  
flavour oscillations or whether a sterile, fourth neutrino is required.

\centerline{\bf Acknowledgements}

S.U. thanks the Theory Group at CERN and the Elementary Particle
Physics group at Universit\"{a}t Mainz for their hospitality when
this work was started. G.B. thanks the Theory Group at Fermilab
for their hospitality where a part of this work was done. We thank
John Ellis for a critical reading of the manuscript. M.N. and
S.U. thank the Department of Science and Technology, Government 
of India, for financial support under project SP/S2/K-13/97.

\newpage

\begin{figure}
\input{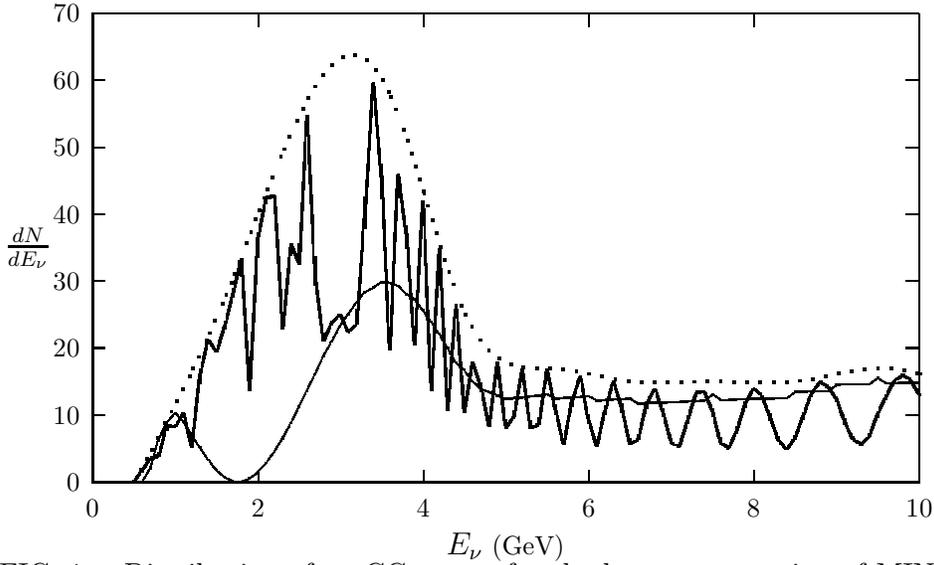}
\caption{
Distribution of $\nu_\mu$ CC events for the low energy option 
of MINOS: Prediction from the
 SB scheme (thick line), for
no oscillations (dots), from the standard
scheme (thin line). The plot from the SB scheme is drawn for
$\Delta_{32} = 0.3$ eV$^2$, $\Delta_{21} = 3 \times
10^{-4}$ eV$^2$, $\theta_{12} = 35^\circ$, $\theta_{13} = 9^\circ$ 
and $\theta_{23} = 27^\circ$. The plot from the standard scheme is 
drawn for $\Delta m^2_{atm} = 3 \times 10^{-3}$ eV$^2$, 
$\theta_{atm} = 45^\circ$.}
\end{figure}

\begin{figure}
\input{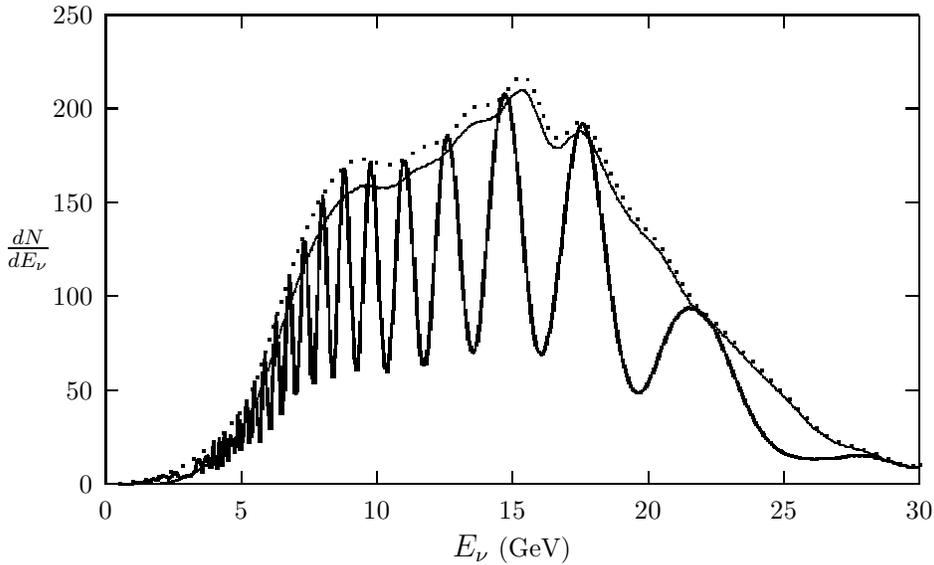}
\caption{Same as Fig.~1, but for the high energy option of 
MINOS.} 
\end{figure}

\begin{figure}
\input{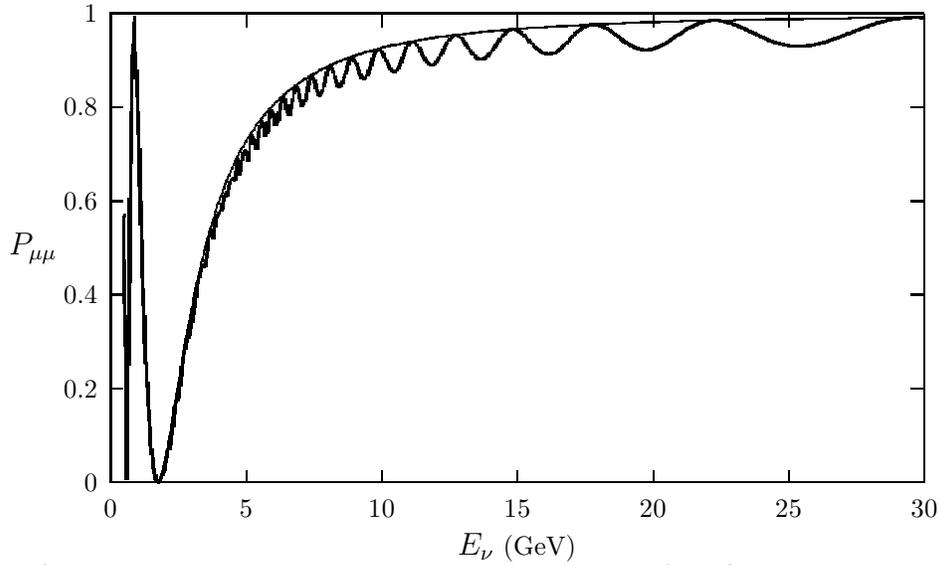}
\caption{Muon neutrino survival probability in the four-flavour
oscillation scheme (thick line) and the standard scheme (thin line).
Both plots are drawn for $\Delta m^2_{atm} = 3 \times 10^{-3}$ 
eV$^2$ and $\theta_{atm} = 45^\circ$. For the four-flavour plot,
we took $\Delta m^2_{LSND} = 0.3$ eV$^2$ and $\sin^2 2    
\theta_{LSND} = 0.1$.} 
\end{figure}

\end{document}